\documentclass[%
 reprint,
 amsmath,amssymb,
 aps,
]{revtex4-1}

\usepackage{graphicx}% Include figure files
\usepackage{dcolumn}% Align table columns on decimal point
\usepackage{bm}% bold math
\begin{document}

\preprint{APS/123-QED}

\title{Greybody factor for black holes in dRGT massive gravity}% Force line breaks with \\
%\thanks{A footnote to the article title}

\author{Petarpa Boonserm}
 \email{petarpa.boonserm@gmail.com}
\affiliation{%
Department of Mathematics and Computer Science, Faculty of Science, Chulalongkorn University, Bangkok 10330, Thailand
}%

\author{Tritos Ngampitipan}
\email{tritos.ngampitipan@gmail.com}
\affiliation{
Faculty of Science, Chandrakasem Rajabhat University, Bangkok 10900, Thailand
}%

\author{Pitayuth Wongjun}
\email{pitbaa@gmail.com}
\affiliation{%
The institute for fundamental study, Naresuan University, Phitsanulok 65000, Thailand
}%

\affiliation{%
Thailand Center of Excellence in Physics, Ministry of Education, Bangkok 10400, Thailand
}%

\date{\today}% It is always \today, today,

\begin{abstract}
In general relativity, greybody factor is a quantity related to the quantum nature of a black hole. A high value of greybody factor indicates a high probability that Hawking radiation can reach infinity. Although general relativity is correct and has been successful in describing many phenomena, there are some questions that general relativity cannot answer. Therefore, general relativity is often modified to attain answers. One of the modifications is the `massive gravity'. The viable model of the massive gravity theory belongs to de Rham, Gabadadze and Tolley (dRGT). In this paper, we calculate the gravitational potential for the de Sitter black hole and for the dRGT black hole. We also derive the rigorous bound on the greybody factor for the de Sitter black hole and the dRGT black hole. It is found that the structure of potentials determines how much the rigorous bound on the greybody factor should be. That is, the higher the potential, the lesser the bound on the greybody factor will be. Moreover, we compare the greybody factor derived from the rigorous bound with the greybody factor derived from the matching technique. The result shows that the rigorous bound is a true lower bound because it is less than the greybody factor obtained from the matching technique.
\begin{description}
\item[Usage]
Secondary publications and information retrieval purposes.
\item[PACS numbers]
 04.62.+v, 04.70.Bw, 04.70.Dy
\item[Structure]
You may use the \texttt{description} environment to structure your abstract;
use the optional argument of the \verb+\item+ command to give the category of each item.
\end{description}
\end{abstract}

\maketitle

%\tableofcontents

\section{\label{sec:level1}Introduction}

General relativity was formulated in 1915. It succeeded in offering profound insights into gravity. Over time, it was verified by many observations. However, numerous cosmological questions remain unanswered such as the hierarchy problem, the cosmological constant problem, and the current acceleration of the Universe. This shows that despite its correctness, general relativity is not a final theory. In the language of particle physics, general relativity is the theory of a massless spin-2 particle \cite{Hinterbichler}. One can generalize general relativity into the massive spin-2 particle theory, which gives mass to a massless spin-2 particle. The most successful theory of massive gravity is known by the de Rham-Gabadadze-Tolley (dRGT) model \cite{dRGT, dRGT2}.

Regarding the cosmological solutions in the dRGT massive gravity theory, even though all the solutions cannot provide a viable cosmological model, for example, the solutions do not admit flat-FLRW metric \cite{D'Amico:2011jj,Gumrukcuoglu:2011ew} or the model encounters instabilities \cite{Gumrukcuoglu:2013nza, D'Amico:2013kya,Chullaphan:2015ija}, a class of solutions can provide a viable cosmological model \cite{DeFelice:2013tsa,DeFelice:2013dua, Tannukij:2015wmn}. The solutions for the dRGT massive gravity are not only investigated in a cosmological background, but also in a spherically symmetric background \cite{Babichev:2015xha}. For spherically symmetric solutions, the black hole solutions have been investigated in both analytical \cite{Volkov, Tasinato, Ghosh:2015cva, Salam, Reentrant, QNM, Hendi, GBInfeld, BTZ, Kanti} and non-analytical \cite{Koyama, Koyama2} forms, depending on the fiducial metric form. However, they still share the same property, which is represented as an asymptotic AdS/dS behavior.

A black hole can emit thermal radiation if the quantum effects are considered. This thermal radiation is known as Hawking radiation \cite{Hawking}. Hawking radiation propagates on spacetime, which is curved by the black hole. The curvature of spacetime acts as a gravitational potential. Therefore, Hawking radiation is scattered from this potential. One part of the Hawking radiation is reflected back into the black hole, while the other part is transmitted to spatial infinity. The transmission probability in this context is also known as the greybody factor.

There are many methods to calculate the greybody factor. For example, one can obtain an approximate greybody factor using the matching technique \cite{Fernando, Kim, Escobedo}. If the gravitational potential is high enough, one can use the WKB approximation to derive the greybody factor \cite{Parikh, Fleming, Lange}. Other than approximation, the greybody factor can also be obtained using the rigorous bound \cite{1D, Bogo, phd}. The bound can give a qualitative description of a black hole.

In this work, we investigate the greybody factor using the analytical black hole solution in dRGT massive gravity. In Section \ref{blackground}, the structure of the horizons of the solution is analyzed in order to generate a suitable form for the analysis of the properties of the greybody factor. In Section \ref{sec:potential}, the properties of the gravitational potential are investigated for both the de Sitter black hole and dRGT black hole. The height of their potentials are determined by the parameters of the model. In Section \ref{rigorous}, we derive the rigorous bound on the greybody factor, and the reflection probability for the de Sitter black hole and the dRGT black hole. The value of the rigorous bound on the greybody factor corresponds to the structure of potentials. In addition, the effects of the graviton mass, cosmological constant and angular momentum quantum number on the greybody factors will also be explored. Finally, concluding remarks are provided in Section \ref{conclu}.

\section{\label{blackground}dRGT black hole background}

In this section, we will review the main concept of the dRGT massive gravity theory in the following manner \cite{Ghosh:2015cva}; the analytical solution of the modified Einstein equation due to the graviton mass is first presented, and then the structure of the horizons of the black hole in the dRGT massive gravity theory is investigated. The theory of the dRGT massive gravity is a covariant non-linear theory of massive gravity, which is ghost free in the decoupling limit to all orders. The action of the dRGT massive gravity model in four-dimensional spacetime can be expressed as
\begin{eqnarray}
S &=& \int d^4 x \sqrt{-g} \left[\frac{M_p^2}{2} R[g] + m^2_g (\mathcal{L}_2[g,f] +\alpha_3\mathcal{L}_3[g,f]\right.\nonumber\\
  &&  \left. + \alpha_4\mathcal{L}_4[g,f] )\right], \label{eqn:action}
\end{eqnarray}
where $R$ is a Ricci scalar corresponding to a physical metric $g_{\mu\nu}$, $m^2_g$ to the square of the graviton mass, with $\mathcal{L}_i$s representing the interactions of the $i$th order of the massive graviton. In particular, those interactions of the massive graviton are constructed from two kinds of metrics and can be expressed as follows,
\begin{eqnarray}
\mathcal{L}_2[g,f] &=& \frac{1}{2}\left([\mathcal{K}]^2-[\mathcal{K}^2]\right), \label{L2} \\
\mathcal{L}_3[g,f] &=& \frac{1}{3!}\left([\mathcal{K}]^3-3[\mathcal{K}][\mathcal{K}^2]+2[\mathcal{K}^3]\right), \label{L3}  \\
\mathcal{L}_4[g,f] &=& \frac{1}{4!}\left([\mathcal{K}]^4 - 6[\mathcal{K}]^2[\mathcal{K}^2] + 3[\mathcal{K}^2]^2 + 8[\mathcal{K}][\mathcal{K}^3]\right.\nonumber\\
                   &&  \left. -6[\mathcal{K}^4]\right), \label{L4}
\end{eqnarray}
where the rectangular brackets denote the traces. The tensor $\mathcal{K}_{\mu\nu}$ is constructed from the physical metric $g_{\mu\nu}$ and the fiducial metric $f_{\mu\nu}$ as
\begin{align}
\mathcal{K}^{\mu}_{\;\nu} = \delta^\mu_{\;\nu} - \left(\sqrt{g^{-1}f}\right)^{\mu}_{\;\nu},
\end{align}
where the square roots of those tensors are defined so that $\sqrt{g^{-1}f}^{\mu}_{\;\rho} \sqrt{g^{-1}f}^{\rho}_{\;\nu} = \left(g^{-1}f\right)^{\mu}_{\;\nu}$. The fiducial metric is chosen as \cite{Volkov}
\begin{eqnarray}\label{fiducial metric}
f_{\mu\nu}=\text{diag}(0,0,c^2  ,c^2 \sin^2\theta),
\end{eqnarray}
where $c$ is a constant. The static and spherically symmetric black hole solution satisfying this theory can be written as \cite{Ghosh:2015cva}
\begin{equation}
ds^{2} = -f(r)dt^{2} + \frac{dr^{2}}{f(r)} + r^{2}d\Omega^{2},
\end{equation}
where
\begin{equation}
f(r) = 1 - \frac{2M}{r} + \frac{\Lambda}{3}r^{2} + \gamma r + \zeta,\label{fr}
\end{equation}
$d\Omega^{2} = d\theta^2 + \sin^2\theta d\phi^2$, and $M$ is an integration constant related to the mass of the black hole. The parameters above can be written in terms of the original parameters as
\begin{subequations}
\begin{eqnarray}
\Lambda&=&3 m^2_g \left(1+\alpha+\beta\right),\\
\gamma&=&-c m^2_g \left(1+2\alpha+3\beta\right),\\
\zeta&=&c^2 m^2_g \left(\alpha+3\beta\right),
\end{eqnarray} \label{pardef}
\end{subequations}
and
\begin{eqnarray}\label{alphabeta}
 \alpha_3 = \frac{\alpha-1}{3}~,~~\alpha_4 =
\frac{\beta}{4}+\frac{1-\alpha}{12}.
\end{eqnarray}
This solution contains various kinds of black hole solutions found in literature. If $m_{g} = 0$, the Schwarzschild solution is recovered. In the case of  $c = 0$, the solution reduces to the de Sitter solution for $1 + \alpha + \beta < 0$ and reduces to the anti-de Sitter solution for $1 + \alpha + \beta > 0$. Moreover, the global monopole solution can be obtained by setting $1+2\alpha+3\beta = 0$. Note that the last term, the constant potential $\zeta$, corresponds to the global monopole term. A global monopole usually comes from a topological defect in high energy physics of the early universe resulting from a gauge-symmetry breaking \cite{bv,Huang:2014oga,Tamaki:2003kv}. However, in this solution, the global monopole is contributed via the graviton mass. Note that the linear term $\gamma r$ is a characteristic term of this solution, distinguished from other solutions found in literature. Next, we will consider the structure of the horizons of this solution. Since the solution is an asymptotical AdS/dS solution, we first consider the structure of the AdS/dS solution and then investigate the structure of the horizons of the solution in the dRGT massive gravity theory.

\subsection{Horizon structure for AdS/dS-like solutions}
\begin{figure}[h!]
\begin{center}
\includegraphics[scale=0.6]{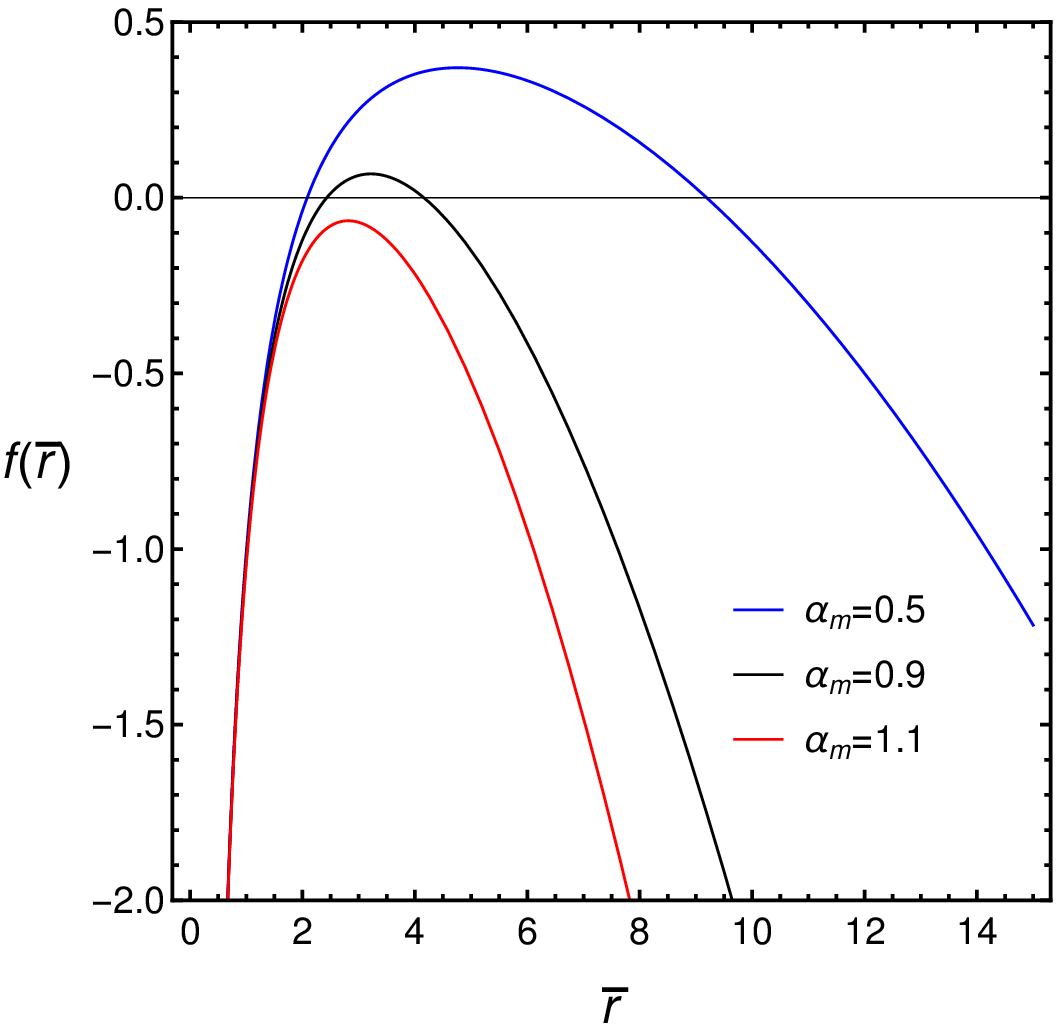}\qquad
\includegraphics[scale=0.6]{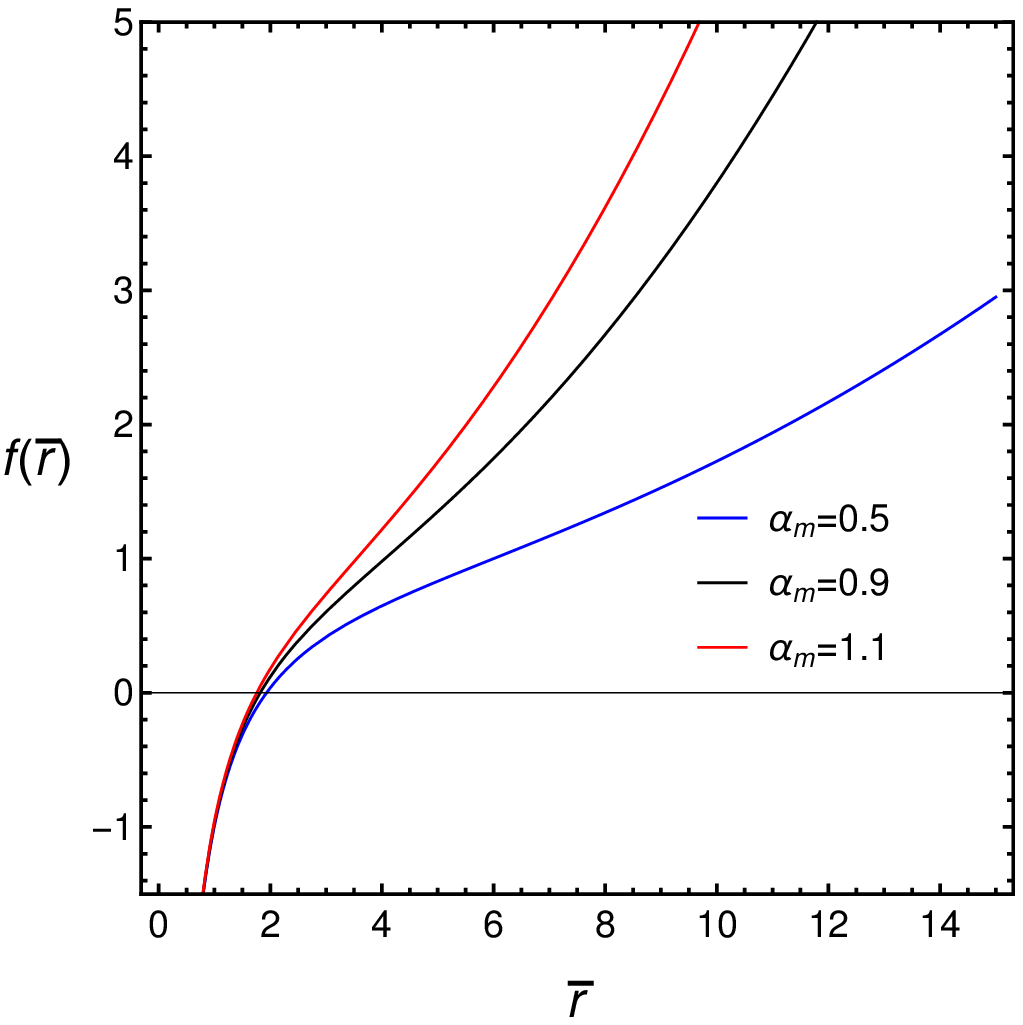}
\end{center}
{\caption{The left panel shows the horizon structure of the de Sitter solution for the value of $\alpha_m$ as $\alpha_m = 0.5$ (Blue line), $\alpha_m = 0.9$ (Black line) and $\alpha_m = 1.1$ (Red line). The right panel shows the horizon structure of the AdS solution for the value of $\alpha_m$ as $\alpha_m = 0.5$ (Blue  line), $\alpha_m = 0.9$ (Black line) and $\alpha_m = 1.1$ (Red line).}\label{fig:dSAdShorizon}}
\end{figure}

It is important to note that one can choose $c =0$. This corresponds to trivial solutions since the interacting terms (or graviton mass) become constant, which is inferred from $\mathcal{K}^{\mu}_{\;\nu} = \delta^\mu_{\;\nu}$. Therefore, the action in Eq. (\ref{eqn:action}) becomes the Einstein-Hilbert action with cosmological constant. In order to investigate the structure of the horizon, let us first consider a simple case where $c=0$. As a result, the function in the metric solution becomes
\begin{equation}
f(r) = 1 - \frac{2M}{r} + \frac{\Lambda}{3}r^{2}. \label{dS-sol}
\end{equation}
From this function, one can see that $f(r) \rightarrow -\infty$ where $r \rightarrow 0$. In order to have two horizons, $f$ must be increased and then decreased where $r$ is increased. This means that $f(r) \rightarrow -\infty$ again when $r\rightarrow \infty$. Therefore, in order to obtain two horizons, $\Lambda$ must be negative. This corresponds to the de Sitter (dS) spacetime, while in the case of anti-de Sitter (AdS) spacetime, $\Lambda > 0$, there exists only one horizon. Now let us find the conditions for which there are two horizons for the de Sitter spacetime, where $\Lambda < 0$. If two horizons exist, the maximum value of $f$ must be positive. The maximum point of $f$ can be found by solving $f' = 0$. As a result, the maximum point is
\begin{equation}
r_m = \left(-\frac{3 M}{\Lambda}\right)^{1/3}.
\end{equation}
Substituting this radius into $f(r)$ in Eq. (\ref{dS-sol}), the maximum value of $f$ can be written as
\begin{equation}
f(r_m) =  \left(-\frac{3 M}{\Lambda}\right)^{1/3} - 3 M.
\end{equation}
By requiring $f(r_m)  > 0$, the condition for having two horizons can be written as
\begin{equation}
-\frac{1}{9M^2} < \Lambda < 0.
\end{equation}
In order to parameterize the solution properly, let us define a dimensionless parameter as
\begin{equation}
\alpha_m^2 = -9\Lambda M^2,
\end{equation}
where $0<\alpha_m^2<1$. By using the dimensionless variable $\bar{r} = r/M$, function $f$ can be rewritten as
\begin{equation}
f(\bar{r}) = 1-\frac{2}{\bar{r}} -\frac{\alpha_m^2}{27} \bar{r}^2.\label{fdS}
\end{equation}
In order to find the horizon, one has to solve the cubic equation;
\begin{equation}
\bar{r}^3-\frac{27}{\alpha_m^2}\bar{r} +\frac{54}{\alpha_m^2} = 0.
\end{equation}
This cubic equation is known as the depressed cubic equation, and the solution can be expressed as
\begin{equation}
\bar{r} = \frac{6}{\alpha_m} \cos\Bigg(\frac{1}{3} \cos^{-1}\Big(-\alpha_{m}\Big) -\frac{2 \pi k}{3} \Bigg),\label{horizon}
\end{equation}
where $k = 0,1,2$ for the three distinguished solutions. Since $0<\alpha_m < 1$, one can expand the sinusoidal function and then keep only the significant contributions. As a result, for $k = 2$, $\bar{r}$ is negative, and for $k = 1$ and $k = 0$, the solutions can be respectively approximated as
\begin{equation}
\bar{r}_1 \sim 2, \,\,\,\,\,\,\,\ \bar{r}_2 \sim \frac{3 \sqrt{3}}{\alpha_m} -1.
\end{equation}
Note that these solutions are well approximated when $\alpha_m \ll 1$. Actually, this approximation can be realized to satisfy the cosmological solution in which the universe expands with acceleration, since the observed value of $\Lambda$ is very small compared to the black hole mass. The behavior of the horizon with various values of $\alpha_m$ is shown in the left panel of Fig. \ref{fig:dSAdShorizon}.

For the AdS solution, $\Lambda$ is positive. Therefore, one can find the solution of the horizon by rewriting $\alpha_m$ as $\alpha_m = 9\Lambda M^2$. By following the same steps from the de Sitter case, the horizon in the AdS case can be written as
\begin{equation}
\bar{r} = \frac{6}{\alpha_m} \sinh\Bigg(\frac{1}{3} \sinh^{-1}\Big(\alpha_{m}\Big) \Bigg).
\end{equation}
As we have discussed above, there exists only one horizon for the AdS solution. This is valid for all values of $\alpha_m$ as show in the right panel of Fig. \ref{fig:dSAdShorizon}.

\subsection{\label{sec:H-structure}Horizon structure for the dRGT massive gravity solutions}
For the complete massive gravity solution, it is significantly difficult and complicated to find the horizon analytically. One of the conditions for having three horizons is that $\Lambda > 0$. Therefore, we can separate our consideration into two classes; the asymptotic AdS solutions for $\Lambda > 0$ and the asymptotic de Sitter solution for $\Lambda < 0$ \cite{Kodama}.

For the general solutions of the dRGT massive gravity, the dimension-length parameter $c$ is not set to be zero. This means that we have to introduce a scale to the theory. It is useful to work out the solution using dimensionless variable, $\tilde{r} = r/c$, and then find out what scale $c$ would assume. As a result, function $f$ can be written in terms of a dimensionless variable as
\begin{equation}
f(\tilde{r}) = 1 - \frac{2\tilde{M}}{\tilde{r}} + \alpha_g \left(c_2\tilde{r}^2 -c_1 \tilde{r}+c_0\right),\label{fdRGT}
\end{equation}
where
\begin{eqnarray}
\tilde{M}&=& \frac{M}{c},\,\,\,\,\,\, \alpha_g = m^2_gc^2,\,\,\,\,\,\, c_0 = \alpha+3\beta,\nonumber\\
c_1&=&1 +2\alpha +3\beta,\,\,\,\,\,\, c_2 = 1+\alpha+\beta.
\end{eqnarray}
From this equation, it is sufficient to figure out that the scale of $c$ takes place at $\tilde{M} \sim \alpha_g$. Therefore, one can choose the parameter $c$ as
\begin{eqnarray}
c = r_V = \left(\frac{M}{m_g^2}\right)^{1/3}.
\end{eqnarray}
This radius is well known as the Vainshtein radius \cite{Vainshtein:1972sx, Arraut}. The theory in which $r < r_V$ will approach GR, while the theory in which  $r > r_V$, the modification of GR will be active. The horizons can be found by solving for the solution of $r$ through the equation
\begin{equation}
\alpha_g c_2 \tilde{r}^3 -\alpha_g c_1 \tilde{r}^2+(\alpha_g c_0+1)\tilde{r} - 2 \tilde{M} =0.
\end{equation}
In order to find the conditions for having three horizons for the AdS case and two horizons for the de Sitter case, let us consider the extremum points of the function $f$ using the equation $f' = 0$ or
\begin{equation}
2 \alpha_g c_2 \tilde{r}^3 -\alpha_g c_1 \tilde{r}^2- 2 \tilde{M} =0.
\end{equation}
This is the cubic equation. We can solve it by changing the variables to obtain the depressed cubic equation and then analyze the general solution to find the condition for having two real positive roots for the AdS case and one real positive root for the de Sitter case. As a result, one can constrain our consideration to case $c_1 = 3 (4c^2_2)^{1/3}$. Choosing this condition will guarantee one real positive root for the de Sitter case as $\tilde{r}_{dS} = (- 2 c_2)^{-1/3}$ and two real positive roots for the AdS case as  $\tilde{r}_{AdS1} = (2 c_2)^{-1/3}$ and $\tilde{r}_{AdS2} = (1+\sqrt{3})(2 c_2)^{-1/3}$.

\begin{figure}[h!]
\begin{center}
\includegraphics[scale=0.6]{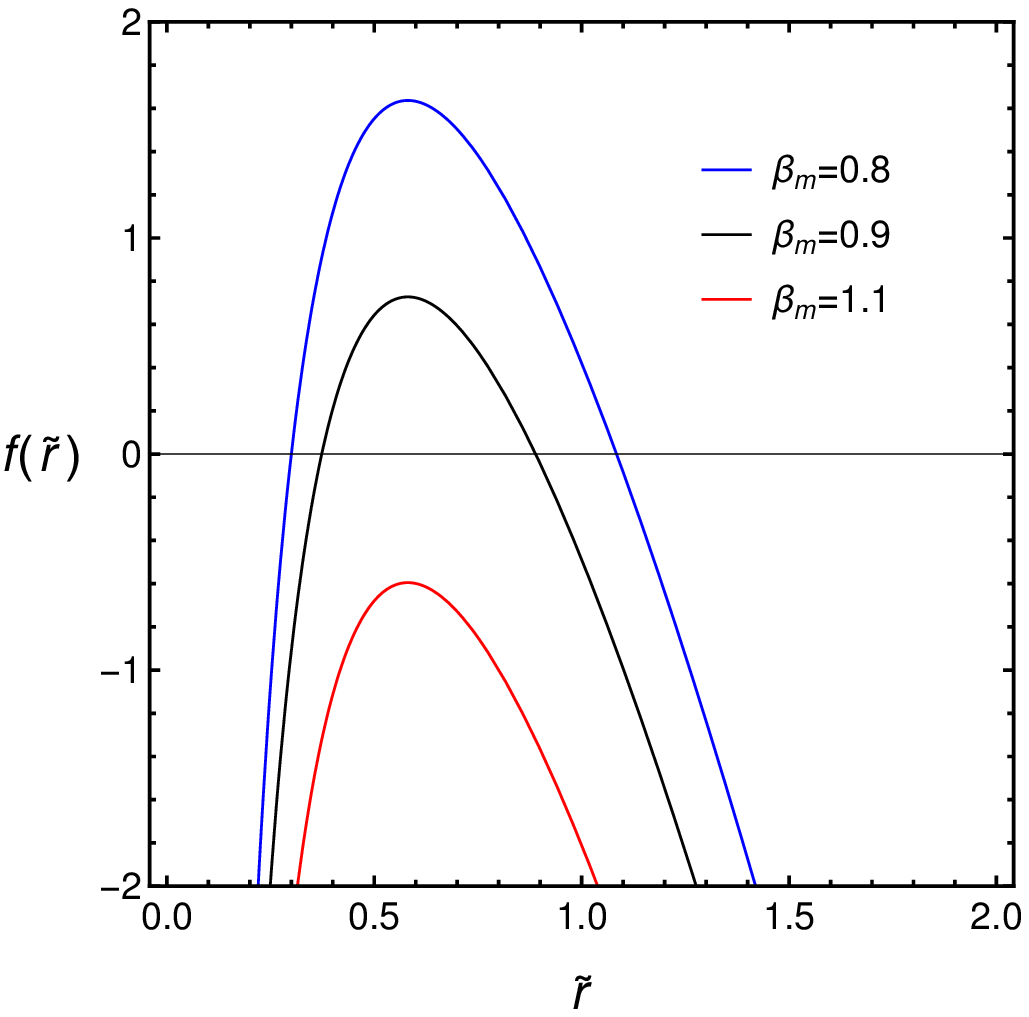}\qquad
\includegraphics[scale=0.6]{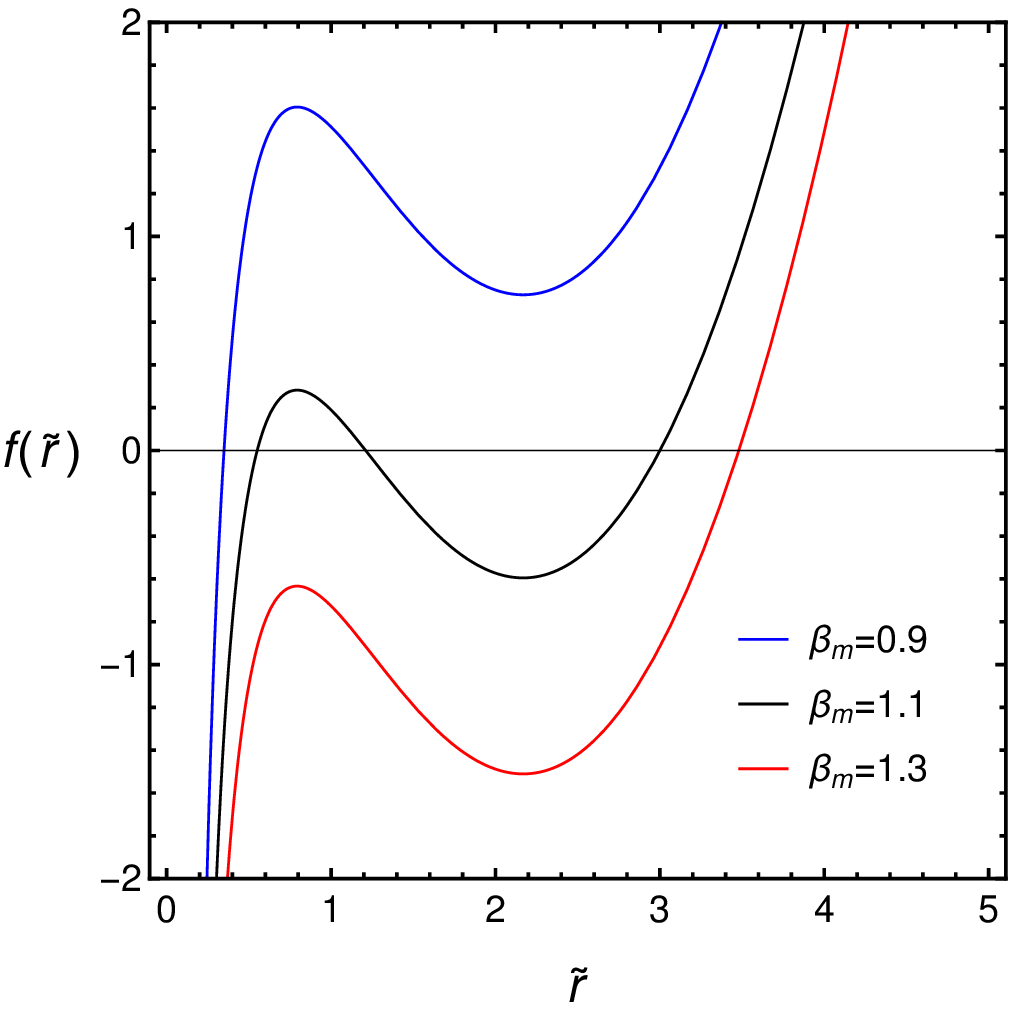}
\end{center}
{\caption{The left panel shows the horizon structure of the asymptotic de Sitter solution in dRGT massive gravity for the value of $\beta_m$ as $\beta_m = 0.8$ (Blue line), $\beta_m = 0.9$ (Black line) and $\beta_m = 1.1$ (Red line). The right panel shows the horizon structure of the asymptotic AdS solution in dRGT massive gravity for the value of $\beta_m$ as $\beta_m = 0.9$ (Blue line), $\beta_m = 1.1$ (Black line) and $\alpha_m = 1.3$ (Red line). We set parameters as $M=1$, $\alpha_g = 1$ and $c_2 = -1$ for the asymptotic de Sitter solution and $c_2 = 1$ for the asymptotic AdS solution.}\label{fig:MGdSAdShorizon}}
\end{figure}

For the asymptotic de Sitter solutions, $f$ at the extremum point can be written as
\begin{equation}
f(\tilde{r}_{dS}) = 1 + c_0 \alpha_g - \frac{9}{\sqrt{3}} \alpha_g \left(-2c_2\right)^{1/3}.
\end{equation}
In order to have two horizons, $f(\tilde{r}_{dS}) > 0$. We can parameterize the parameter $c_0$ such that
\begin{equation}
c_0 = \frac{9}{\sqrt{3}} \frac{\left(-2c_2\right)^{1/3}}{\beta_m} - \frac{1}{\alpha_g}. \label{c0eq}
\end{equation}
Therefore, the condition for having two horizons in the case of de Sitter-like spacetime is
\begin{equation}
0 < \beta_m < 1.
\end{equation}
By changing the cubic equation into the depressed cubic equation, one can find the two real positive horizons as the real roots of the depressed cubic equation as follows
\begin{eqnarray}
\tilde{r}_{1} &=& \frac{2}{\left(-2c_2\right)^{1/3}}\left[X^{1/2}\cos\left(\frac{1}{3}\sec^{-1}Y\right) - 1\right],\label{rmg1}\\
\tilde{r}_{2} &=& \frac{-2}{\left(-2c_2\right)^{1/3}}\left[X^{1/2}\cos\left(\frac{1}{3}\sec^{-1}Y + \frac{\pi}{3}\right) + 1\right],\nonumber\\\label{rmg2}
\end{eqnarray}
where
\begin{equation}
X = \frac{2\sqrt{3}}{\beta_m} + 4 ~\text{and}~ Y = -\frac{\sqrt{\frac{\sqrt{3}}{\beta_m} + 2}\left(2\sqrt{2}\beta_m + \sqrt{6}\right)}{5\beta_m + 3\sqrt{3}}.
\end{equation}
Our analysis can be checked using the numerical method as shown in the left panel of Fig. \ref{fig:MGdSAdShorizon}. From this figure, one can see that $\beta_m$ can parameterize the existence of two horizons.

Now we consider the asymptotic AdS solutions. The $f$ at the extremum points can be written as
\begin{eqnarray}
f(\tilde{r}_{AdS1}) = 1 + c_0 \alpha_g - \frac{9}{2} \alpha_g \left(2c_2\right)^{1/3},\\
f(\tilde{r}_{AdS2}) = 1 + c_0 \alpha_g - \frac{9}{\sqrt{3}} \alpha_g \left(2c_2\right)^{1/3}.
\end{eqnarray}
In order to have three horizons, we must have $f(\tilde{r}_{AdS1}) > 0$ and $f(\tilde{r}_{AdS2}) < 0$. By using the parameter of $\alpha_0$ in Eq. (\ref{c0eq}), while changing $c_2$ to $-c_2$,  the condition for having three horizons can be written as
\begin{eqnarray}
1 < \beta_m < \frac{2}{\sqrt{3}}.
\end{eqnarray}
By using the same step as done in the asymptotic de Sitter case, the three real positive horizons for the AdS case can be written as
\begin{eqnarray}
\tilde{r}_{1} &=& \frac{2}{\left(2c_2\right)^{1/3}}\left[1 - x^{1/2}\sin\left(\frac{1}{3}\sec ^{-1}y + \frac{\pi}{6}\right)\right],\\
\tilde{r}_{2} &=& \frac{2}{\left(2c_2\right)^{1/3}}\left[1 - x^{1/2}\cos\left(\frac{1}{3}\sec^{-1}y + \frac{\pi}{3}\right)\right],\\
\tilde{r}_3 &=& \frac{2}{\left(2c_2\right)^{1/3}}\left[1 + x^{1/2}\cos\left(\frac{1}{3}\sec^{-1}y\right)\right].
\end{eqnarray}
where
\begin{equation}
x = \frac{4 - 2\sqrt{3}}{\beta_m} ~\text{and}~ y = \frac{\sqrt{6}-2 \sqrt{2} \beta _m}{\left(3 \sqrt{3}-5 \beta _m\right) \sqrt{-\frac{\beta _m}{\sqrt{3}-2 \beta _m}}}.
\end{equation}
The numerical plot for these horizons is shown in the right panel of Fig. \ref{fig:MGdSAdShorizon}. From this figure, one can see that the existence of three horizons satisfy the condition $1< \beta_m < 2/\sqrt{3}$ as we have analyzed. In the next section, we will use the expression for the horizons derived in this section to analyze the properties of the gravitational potential and the greybody factor of the black hole.

\section{\label{sec:potential}Equations of motion of massless scalar field}
\begin{figure}[h]
\begin{center}
\includegraphics[scale=0.6]{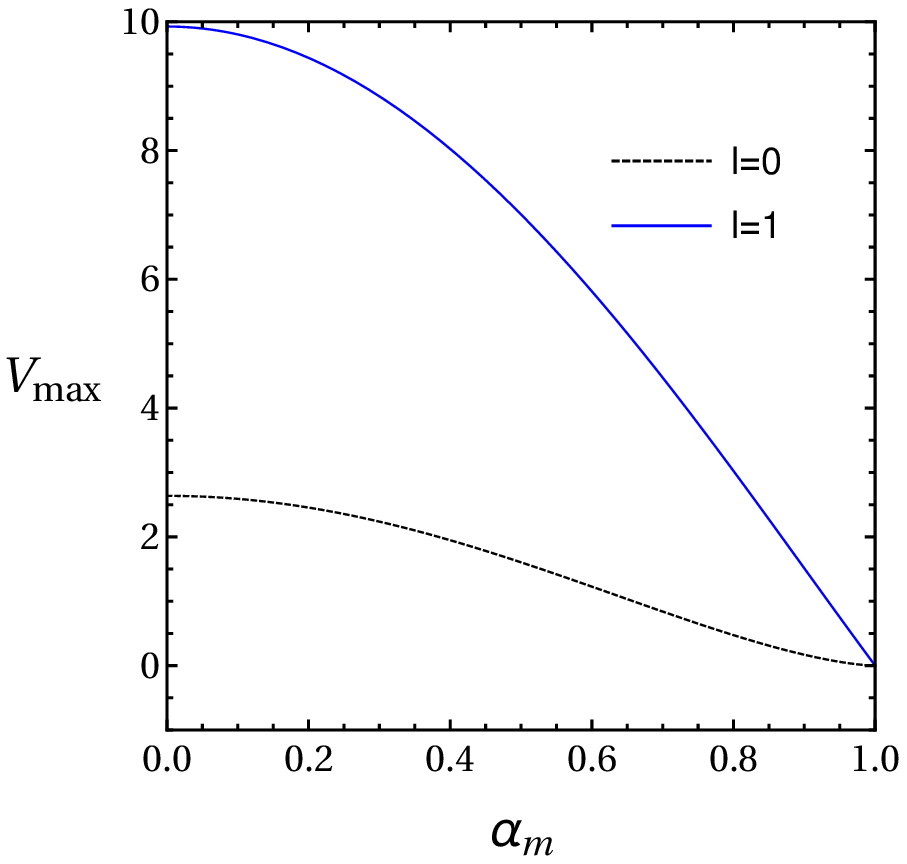}\qquad\qquad
\includegraphics[scale=0.6]{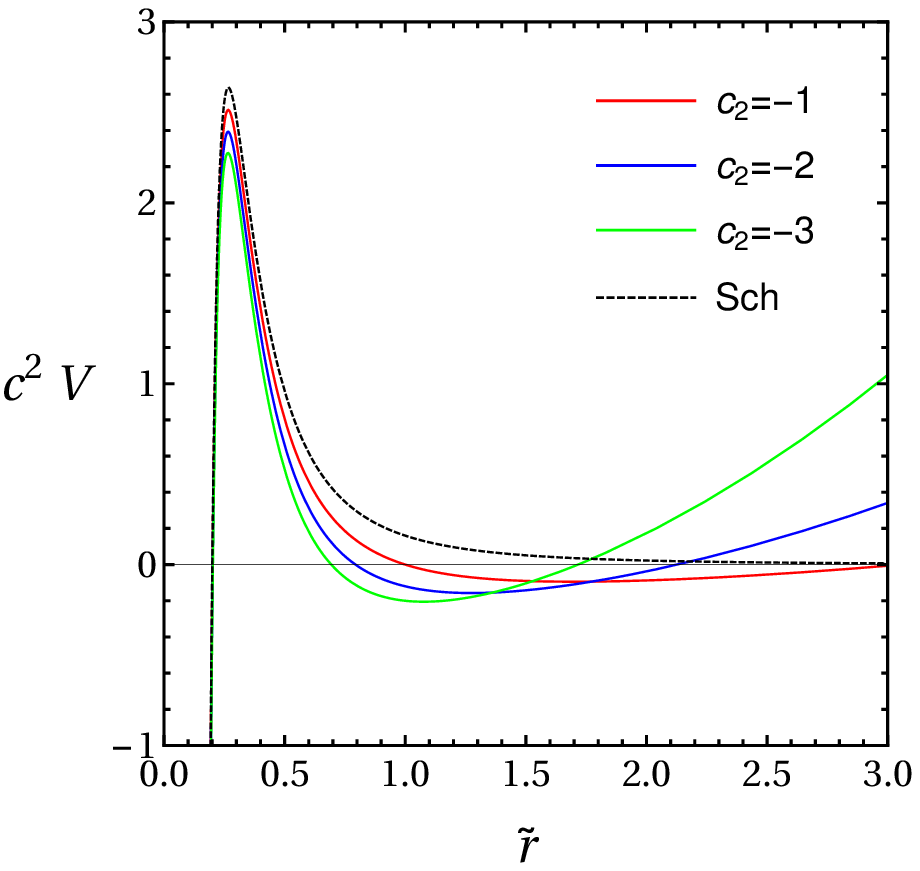}
\end{center}
{\caption{The left panel shows the maximum potential for a de Sitter black hole versus the model parameter $\alpha_m$ with $\ell = 0,\ell = 1$, $\tilde{M}=\alpha_{g}= 0.1$.  The right panel shows  the potential for a de Sitter black hole with different values of $c_2$ compared to the Schwarzschild case; black-dotted line for the Schwarzschild case, red line for $c_2=-1$, blue line for $c_2=-2$ and green line for $c_2 = -3$.}\label{vsch-vds}}
\end{figure}

Classically, nothing can escape a black hole when approaching it. However, when a quantum effect is considered, the black hole can radiate. This radiation is known as Hawking radiation. It is a blackbody spectrum of temperature
\begin{equation}
kT = \frac{\hbar}{4\pi r_{s}},
\end{equation}
where $r_{s}$ is the Schwarzschild radius. In this paper, we assume that Hawking radiation is a massless scalar field. The massless scalar field satisfies the Klein-Gordon equation
\begin{equation}
\frac{1}{\sqrt{-g}}\partial_{\mu}\left(\sqrt{-g}g^{\mu\nu}\partial_{\nu}\Phi\right) = 0.
\end{equation}
We use the spherical coordinates. The solutions to the wave equation in spherical coordinates are of the form
\begin{equation}
\Phi(t, r, \Omega) = e^{i\omega t}\frac{\psi(r)}{r}Y_{\ell m}(\Omega),
\end{equation}
where $Y_{\ell m}(\Omega)$ are spherical harmonics. The Klein-Gordon equation becomes
\begin{eqnarray}
\frac{\omega^{2}r^{2}}{f(r)} + \frac{r}{\psi(r)}\frac{d}{dr}\left[r^{2}f(r)\frac{d}{dr}\left(\frac{\psi(r)}{r}\right)\right] &&\nonumber\\
+ \frac{1}{Y(\Omega)}\left[\frac{1}{\sin\theta}\frac{\partial}{\partial\theta}\left(\sin\theta\frac{\partial Y(\Omega)}{\partial\theta}\right)\right] &&\nonumber\\
+ \frac{1}{\sin^{2}\theta}\frac{1}{Y(\Omega)}\frac{\partial^{2}Y(\Omega)}{\partial\phi^{2}} &=& 0.\label{KG}
\end{eqnarray}
The angular part satisfies
\begin{equation}
\frac{1}{\sin\theta}\frac{\partial}{\partial\theta}\left(\sin\theta\frac{\partial Y(\Omega)}{\partial\theta}\right) + \frac{1}{\sin^{2}\theta}\frac{\partial^{2}Y(\Omega)}{\partial\phi^{2}} = -\ell(\ell + 1)Y(\Omega),
\end{equation}
where $\ell$ is the angular momentum quantum number. Therefore, the Klein-Gordon equation (Eq. (\ref{KG})) is left with the radial part
\begin{equation}
\frac{d^{2}\psi(r)}{dr_{*}^{2}} + \left[\omega^{2} - V(r)\right]\psi(r) = 0,
\end{equation}
where $r_{*}$ is the tortoise coordinate defined by
\begin{equation}
\frac{dr_{*}}{dr} = \frac{1}{f(r)}
\end{equation}
and $V(r)$ is the potential given by
\begin{equation}
V(r) = \frac{\ell(\ell + 1)f(r)}{r^{2}} + \frac{f(r)f'(r)}{r}.\label{poten}
\end{equation}
It can be expressed in terms of $\tilde{r}$ as
\begin{equation}
V(\tilde{r}) = \frac{\ell(\ell + 1)f(\tilde{r})}{c^{2}\tilde{r}^{2}} + \frac{f(\tilde{r})f'(\tilde{r})}{c^{2}\tilde{r}}.\label{potentilde}
\end{equation}
Substituting the function $f(\tilde{r})$ from Eq. (\ref{fdRGT}), we obtain
\begin{eqnarray}
V(\tilde{r}) &=& \left[1 - \frac{2\tilde{M}}{\tilde{r}} + \alpha_g \left(c_2\tilde{r}^2 -c_1 \tilde{r}+c_0\right)\right]\nonumber\\
             &&  \times\left[\frac{\ell(\ell + 1)}{c^{2}\tilde{r}^{2}} + \frac{1}{c^{2}\tilde{r}}\left[\frac{2\tilde{M}}{\tilde{r}^{2}} + \alpha_g \left(2c_2\tilde{r} - c_1\right)\right]\right].\nonumber\\ \label{poten mg}
\end{eqnarray}
From the above equation, we can see that the potential is high when the angular momentum quantum number is large. Qualitatively, the leading contribution to the transmission amplitude comes from the mode $\ell=0$. Therefore, it is sufficient to qualitatively analyze the potential for the case of $\ell=0$. When $\ell = 0$, the first term in Eq. \ref{potentilde} vanishes and the second term  is proportional to $f f'$. Therefore, there are three $\tilde{r}$-intercepts resulting from $f = 0$ and $f' = 0$. This behavior significantly differs from the Schwarzschild case, which has only one $\tilde{r}$-intercept as shown in the right panel of Fig. \ref{vsch-vds}.

For the de Sitter case, the potential can be obtained by setting $c_1=c_0 = 0$. As we have analyzed in Section \ref{sec:H-structure}, it is convenient to change parameter as $\alpha_g c_2 = - \alpha_{m}^{2}/ (27 \tilde{M}^2)$. For this setting, the potential depends only on the parameter $\alpha_m$. In the same strategy as the quantum theory, the shape of the potential controls the transmission amplitude. Therefore, it is worthwhile to consider the maximum value of the de Sitter potential compared to the Schwarzschild potential. As a result, the de Sitter potential for the $\ell=0$ case can be written as
\begin{equation}
c^{2}V_{\text{dS}}(\tilde{r}) = \frac{1}{\tilde{r}}\left(1 - \frac{2\tilde{M}}{\tilde{r}} -\frac{\alpha_m^2}{27 \tilde{M}^2}\tilde{r}^{2}\right)\left(\frac{2\tilde{M}}{\tilde{r}^{2}} -\frac{2\alpha_m^2}{27 \tilde{M}^2}\tilde{r}\right).
\end{equation}
By solving $\tilde{r}_{\max}$ via $V_{\text{dS}}'=0$ and then substituting the solution back into the above equation, one can find the maximum value of the potential depending on only two parameters, $\tilde{M}$ and $\alpha _m$. The expression is significantly lengthy; we do not present it in the current paper. In order to see the effect of the graviton mass or the cosmological constant, one can fix $\tilde{M}$ and then plot this expression via $\alpha _m$ as shown in the left panel of Fig. \ref{vsch-vds}. Note that we also show the result for the $\ell=1$ case in this figure. From this figure, one can see that $V_{\max}$ contributed from the de Sitter black hole is always less than one from the Schwarzschild black hole. Clearly, the cosmological constant plays a role in reducing the local maximum  of the potential. The explicit form of the de Sitter potential is plotted with various values of the cosmological constant as shown in the right panel of Fig. \ref{vsch-vds}. Note that we used the parameter $c_2$ instead of $\alpha_m^2$. This is convenient for comparing the results with one from the case of the dRGT massive gravity. Consequently, it might be expected that the transmission amplitude due to the de Sitter black hole should be greater than one in the Schwarzschild black hole. We will clarify this issue explicitly in the next section.

\begin{figure}[h]
\begin{center}
\includegraphics[scale=0.6]{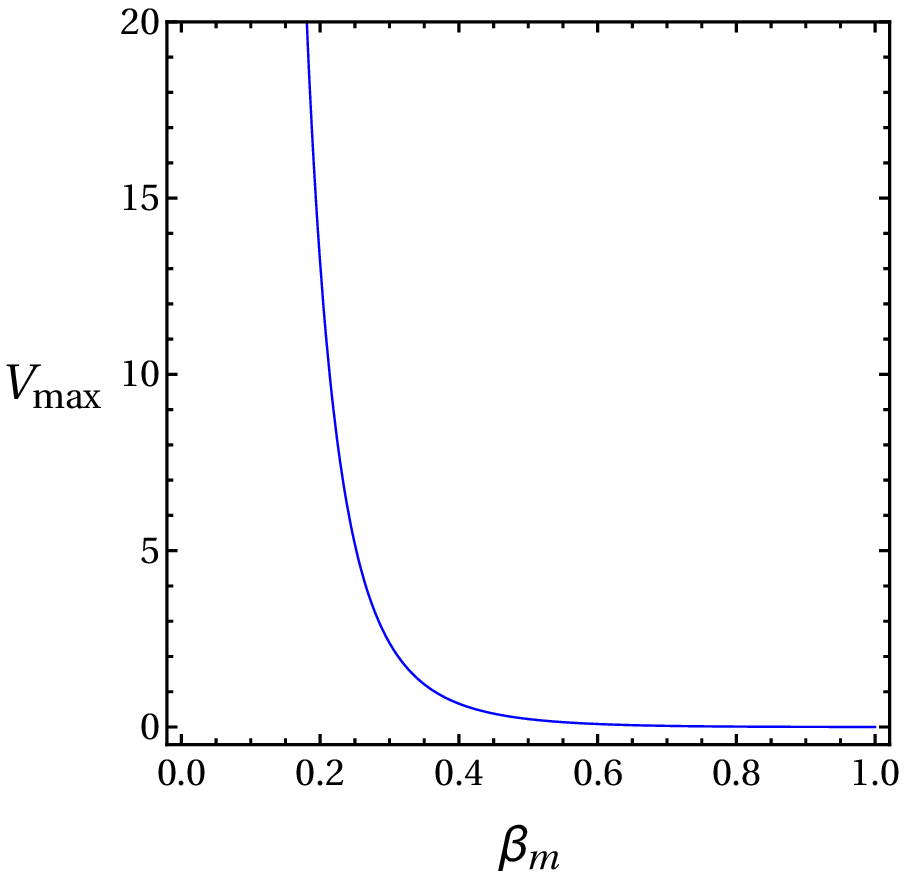}\qquad\qquad
\includegraphics[scale=0.6]{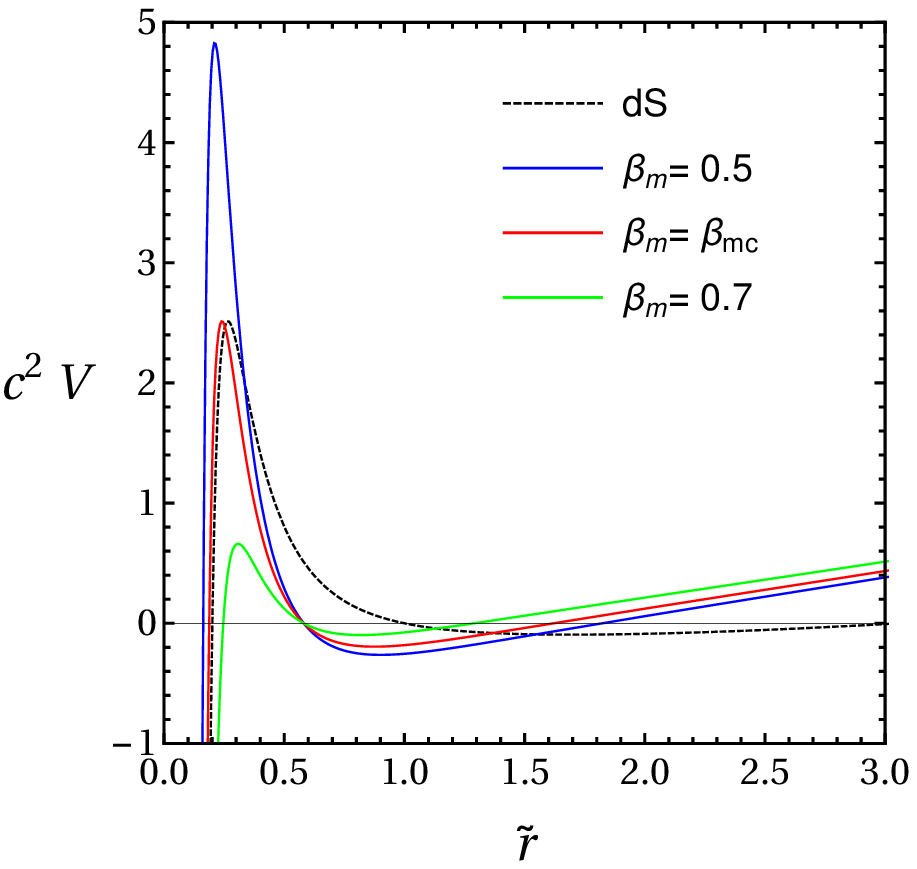}
\end{center}
{\caption{The left panel shows the maximum of dRGT potential with $\ell$ = 0, $\tilde{M}$ = 0.1, $\alpha_{g}$ = 0.1, $c$ = 1, and $c_2=-1$. The right panel shows  the dRGT potential with $\ell$ = 0, $\tilde{M}$ = 0.1, $\alpha_{g}$ = 0.1, $c$ = 1, and $c_2=-1$}\label{pot}}
\end{figure}

The dRGT potential for $\ell = 0$ can explicitly be written  as
\begin{eqnarray}
V_{\text{dRGT}}(\tilde{r}) &=& \frac{1}{c^{2}\tilde{r}}\left[1 - \frac{2\tilde{M}}{\tilde{r}} + \alpha_{g}c_{2}\tilde{r}^{2} - 3\sqrt[3]{4c_{2}^{2}}\alpha_{g}\tilde{r}\right.\nonumber\\
                           &&  \left. + \frac{3\sqrt{3}\alpha_{g}\sqrt[3]{-2c_{2}}}{\beta_{m}} - 1\right]\nonumber\\
                           &&  \times \left(\frac{2\tilde{M}}{\tilde{r}^{2}} + 2\alpha_{g}c_{2}\tilde{r} - 3\sqrt[3]{4c_{2}^{2}}\alpha_{g}\tilde{r}\right).
\end{eqnarray}
By employing the same strategy as used in the de Sitter case, we found that the leading term of $V_{\max}$ is proportional to $V_{\max} \propto 1/\beta_m$. By fixing $c_2 = -1$, we have illustrated the explicit behavior of the peak of the potential in the left panel of Fig. \ref{pot}. We can explicitly see that $V_{\max}$ increases as $\beta_{m}$ decreases. Moreover, the de Sitter potential and the dRGT potential with various values of $\beta_m$ are plotted as shown in the right panel of Fig. \ref{pot}. It shows that both the de Sitter potential and the dRGT potential increase with radial distance from the black hole. After that, they decrease with radial distance to reach the relative lowest point and then turn to increase again.

\begin{figure}[h]
\begin{center}
\includegraphics[scale=0.6]{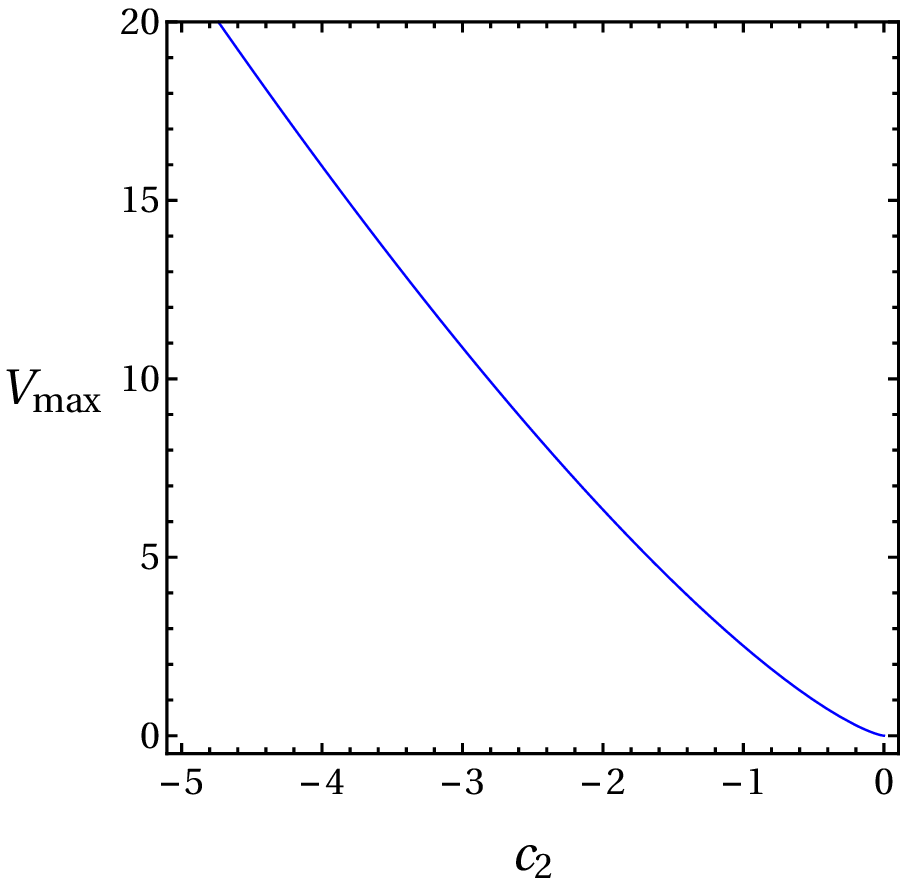}\qquad\qquad
\includegraphics[scale=0.6]{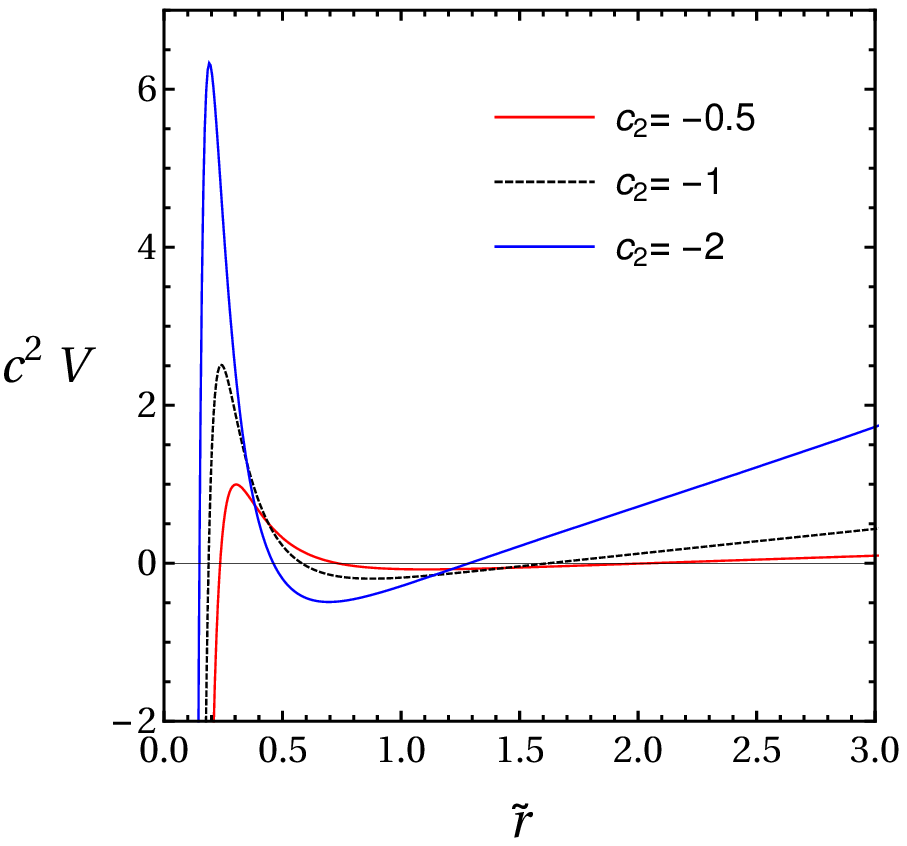}
\end{center}
{\caption{The left panel shows the maximum dRGT potential with $\ell$ = 0, $\tilde{M}$ = 0.1, $\alpha_{g}$ = 0.1, $c$ = 1, and $\beta_{m}$ = 0.565375. The right panel shows the dRGT potential with $\ell$ = 0, $\tilde{M}$ = 0.1, $\alpha_{g}$ = 0.1, $c$ = 1, and $\beta_{m}$ = 0.565375.}\label{c2pot}}
\end{figure}

In \cite{china}, the maximum points of the potentials are not chosen to be equal. We consider this point here. The equality of the maximum points allows us to draw conclusions regarding how high the rigorous bounds on the greybody factors for different types of black holes are. At the highest point, the derivative of the potential is zero. For $\ell = 0$, we obtain
\begin{equation}
V'(\tilde{r}) = \frac{1}{c^{2}}\frac{\tilde{r}f(\tilde{r})f''(\tilde{r}) + \tilde{r}[f'(\tilde{r})]^{2} - f(\tilde{r})f'(\tilde{r})}{\tilde{r}^{2}}.
\end{equation}
The solution of $V'(\tilde{r}) = 0$ is not shown here. We find that the equality of the peak of the potentials occur at $\beta_{m} = 0.565375$. Moreover, the effects of parameter $c_2$  are shown in Fig. \ref{c2pot}.

To see the effect of the parameter $c_2$ on the potential, let us fix $\beta_m = \beta_{mc}$. The peak of the potential is plotted as shown in the left panel of Fig. \ref{c2pot}. The potential is also plotted with various values of $c_2$ as shown in the right panel of Fig. \ref{c2pot}. The parameter $c_2$ characterizes the strength of the graviton mass. Therefore, the graviton mass will enhance the potential in contrast to the effect of the cosmological constant in the de Sitter black hole.

In this section, we explore the behavior of the gravitational potential for both the de Sitter black hole and the dRGT black hole of a massless scalar field. By making a comparison with the potential in the Schwarzschild black hole, we found that the local maximum of the de Sitter potential is always less than one of the Schwarzschild potential. For the dRGT black hole, the local maximum of the potential depend on the model parameters; $\beta_m$ characterizing the existence of two horizon ($0 < \beta_m < 1 $) and $c_2$ characterizing the strength of the graviton mass. In contrast to the de Sitter potential, we found that the local maximum of the dRGT potential will be larger than ones for the Schwarzschild and the de Sitter potential by setting parameter $\beta_m \ll 1$ or $c_2 \ll -1$. In the same fashion as quantum theory, the shape of the potential has an effect on the transmission amplitude or the the greybody factor in this context. We will use the information of the potential to analyze the behavior of the greybody factor in the next section.

\section{\label{rigorous}The rigorous bounds on the greybody factors}
In this section, a greybody factor will be obtained using the rigorous bound \cite{1D, Bogo, phd, Tphd}. The bound can give a qualitative description of a black hole. It is applied to various types of black holes such as a Schwarzschild black hole \cite{Sch}, a non-rotating black hole \cite{non}, a dirty black hole \cite{dirty}, a Kerr-Newman black hole \cite{KN}, a Myers-Perry black hole \cite{MP}, and a dRGT black hole \cite{china}. The rigorous bounds on the greybody factors are given by
\begin{equation}
T \geq \text{sech}^{2}\left(\int_{-\infty}^{\infty}\vartheta dr_{*}\right),
\end{equation}
where
\begin{equation}
\vartheta = \frac{\sqrt{[h'(r_{*})]^{2} + \left[\omega^{2} - V(r_{*}) - h^{2}(r_{*})\right]^{2}}}{2h(r_{*})},
\end{equation}
where $h(r_{*})$ is a positive function satisfying $h(-\infty) = h(\infty) = \omega$. See \cite{1D} for more details. We select $h = \omega$. Therefore,
\begin{equation}
T \geq \text{sech}^{2}\left(\frac{1}{2\omega}\int_{-\infty}^{\infty}|V|dr_{*}\right).\label{bound}
\end{equation}

\subsection{de Sitter black holes}
\begin{figure}[h]
\begin{center}
\includegraphics[scale=0.6]{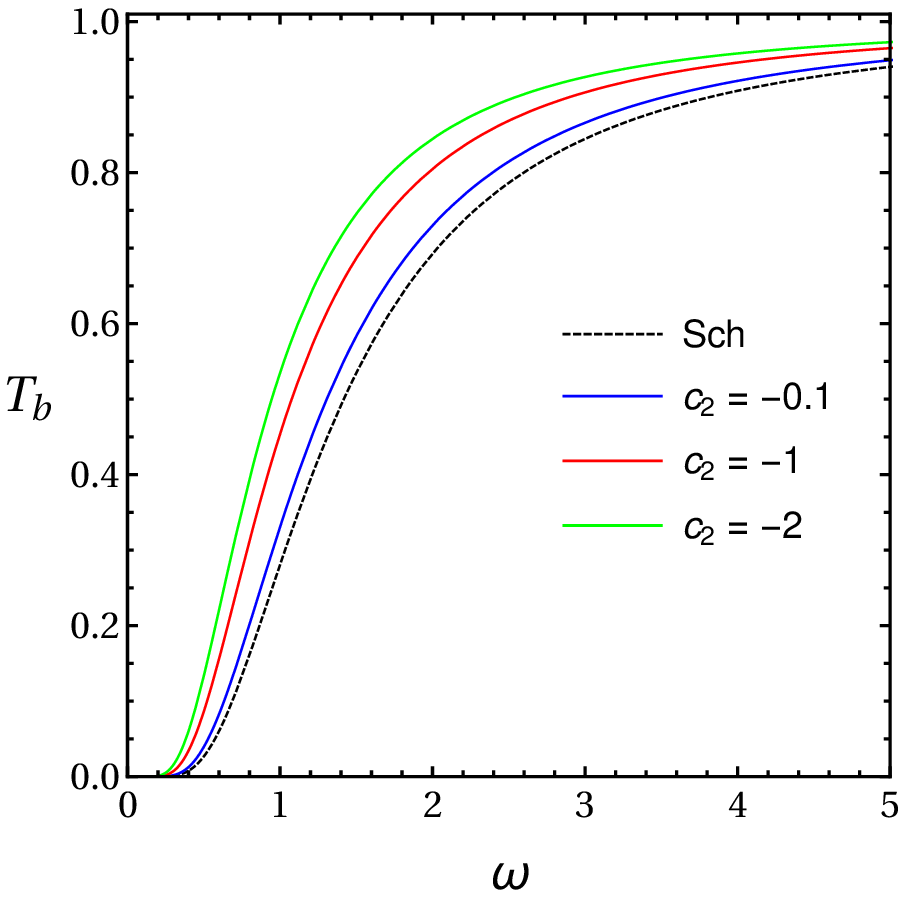}\qquad\qquad
\includegraphics[scale=0.6]{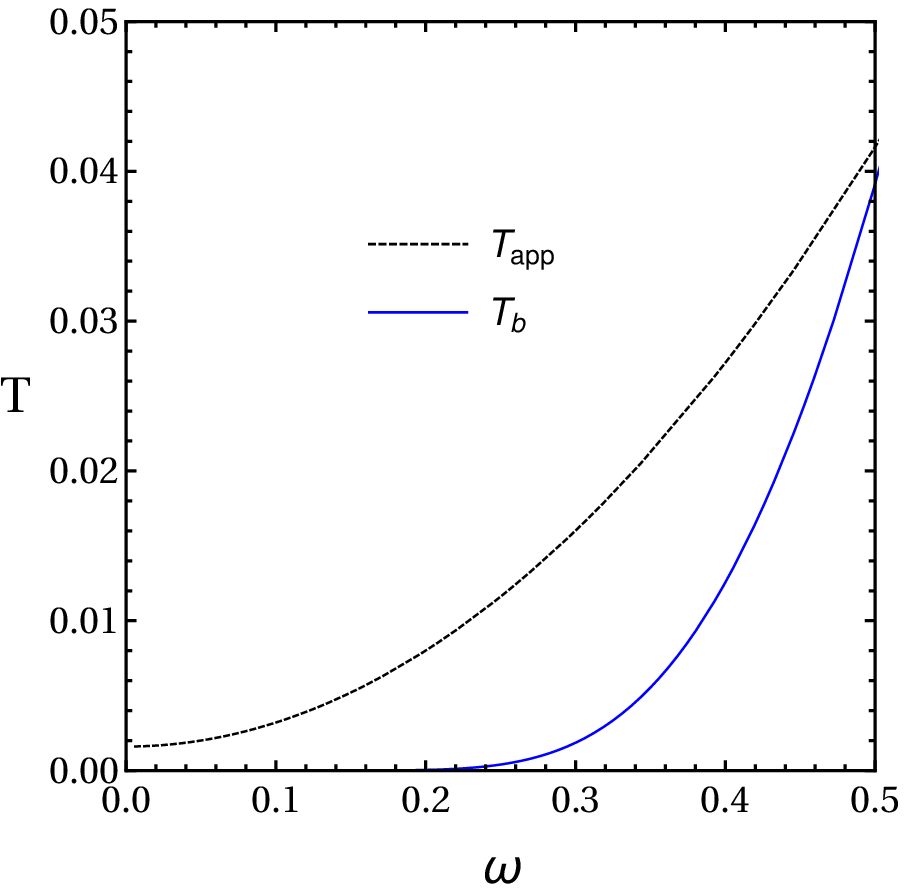}
\end{center}
{\caption{The left panel shows a comparison between the rigorous bound on the greybody factor for a de Sitter black hole and a Schwarzschild black hole with $\ell$ = 0, $c$ = 1, and  $\tilde{M}=\alpha_{g} = 0.1$. The right panel shows a comparison between the rigorous bound on the greybody factor and the approximation with $\ell$ = 0, $c$ = 1, $c_{2}= -1$  and $\tilde{M}=\alpha_{g} = 0.1$.}\label{ds-sch}}
\end{figure}

To obtain the rigorous bound, we use the potential derived in the previous section. For de Sitter black holes, the potential is given by Eq. (\ref{poten mg}), with $c_{1} = c_{0} = 0$. Substituting this potential into Eq. (\ref{bound}), we obtain the rigorous bounds on the greybody factors
\begin{eqnarray}
T \geq T_{b} &=& \text{sech}^{2}\left[\frac{1}{2\omega c}\left\{\ell(\ell + 1)\left(\frac{1}{\tilde{r}_{H}} - \frac{1}{\tilde{R}_{H}}\right)\right.\right.\nonumber\\
             &&  \left.\left. + \tilde{M}\left(\frac{1}{\tilde{r}_{H}^{2}} - \frac{1}{\tilde{R}_{H}^{2}}\right) + 2\alpha_{g}c_{2}\left(\tilde{R}_{H} - \tilde{r}_{H}\right)\right\}\right].\nonumber\\ \label{boundT}
\end{eqnarray}
The rigorous bounds on the reflection probabilities are given by
\begin{eqnarray}
R &\leq& \tanh^{2}\left[\frac{1}{2\omega c}\left\{\ell(\ell + 1)\left(\frac{1}{\tilde{r}_{H}} - \frac{1}{\tilde{R}_{H}}\right)\right.\right.\nonumber\\
  &&     \left.\left. + \tilde{M}\left(\frac{1}{\tilde{r}_{H}^{2}} - \frac{1}{\tilde{R}_{H}^{2}}\right) + 2\alpha_{g}c_{2}\left(\tilde{R}_{H} - \tilde{r}_{H}\right)\right\}\right],
\end{eqnarray}
where, from Eq. (\ref{horizon}), $\tilde{r}_{H}$ and $\tilde{R}_{H}$ are given by
\begin{eqnarray}
\tilde{r}_{H} &=& \frac{6\tilde{M}}{\alpha_m} \cos\Bigg[\frac{1}{3} \cos^{-1}\Big(-\alpha_{m}\Big) - \frac{2\pi}{3} \Bigg]\label{rH}\\
\tilde{R}_{H} &=& \frac{6 \tilde{M}}{\alpha_m} \cos\Bigg[\frac{1}{3} \cos^{-1}\Big(-\alpha_{m}\Big)\Bigg].\label{RH}
\end{eqnarray}
Since $\tilde{r}_H $ and $\tilde{R}_H$ depend on parameters $\tilde{M}$ and $\alpha_m$, the structure of $T_b$ depends on the strength of the cosmological constant through the parameter $\alpha_m$. To see the effect of the cosmological constant qualitatively, let us consider the case $\alpha_m \ll 1$, which means that the effect of the cosmological constant is a correction to the Schwarzschild case. As a result, the horizons can be approximated as
\begin{eqnarray}
\tilde{r}_{H} \sim  2\tilde{M}, \,\,\,\,\,\,\tilde{R}_{H} \sim \left(\frac{3\sqrt{3}}{\alpha_m} - 1\right)\tilde{M}.
\end{eqnarray}
By substituting these results into Eq. \ref{boundT}, the rigorous bound on the greybody factor for a de Sitter black hole can be approximated as
\begin{eqnarray}
T_{b} &=& \text{sech}^{2}\left[\frac{1}{2\omega c}\left\{\frac{\ell(\ell + 1)}{\tilde{r}_H }\left(1- \frac{2|\alpha_m|}{3\sqrt{3}}\right)\right.\right.\nonumber\\
      &&  \left.\left. + \frac{\tilde{M}}{r_H^2}\left(1 - \frac{4\alpha_m^2}{27}\right) - \frac{2\sqrt{3}|\alpha_m|}{9\tilde{M}} \right\}\right].\label{boundT-app}
\end{eqnarray}
From this equation, one can see that if $\alpha_m = 0$, the bound for Schwarzschild is recovered. Moreover, for $\alpha_m \neq 0$, the cosmological constant provides a negative correction to the Schwarzschild bound. Therefore, the greybody factor for the de Sitter black hole is greater than one for the Schwarzschild black hole. This is also consistent with the behavior of the potential, since the local maximum of the potential in the de Sitter black hole is always less than one in the Schwarzschild black hole. To confirm this result, we also used a numerical method to show that $T_{b(dS)} \geq T_{b(Sch)}$ by plotting $T_b$ with various values of $c_2$ as illustrated in the left panel of Fig. \ref{ds-sch}.

The rigorous bound on the greybody factor is useful in any problem, especially in qualitative work. Moreover, the rigorous bound is accurate and its method of derivation is simpler than any other method such as the approximation derived from the matching technique. To see this, let us compare the rigorous bound with the matching technique. The analytical approximation from the matching technique in the low frequency limit for $\ell = 0$ is given by \cite{Harmark, Dong}
\begin{equation}
T_{\text{app}} = 4(\kappa r_{H})^{2}\left(1 + \frac{\omega^{2}}{\kappa^{2}}\right) = 4(\kappa c\tilde{r}_{H})^{2}\left(1 + \frac{\omega^{2}}{\kappa^{2}}\right),
\end{equation}
where $\kappa^{2} = -\alpha_{g}c_{2}/c^{2}$. The rigorous bound on the greybody factor and the approximation are plotted as shown in the right panel of Fig. \ref{ds-sch}. The graph shows that the rigorous bound is less than the approximation, which satisfies the inequality (\ref{boundT}).

\subsection{dRGT black holes}
\begin{figure}[h]
\begin{center}
\includegraphics[scale=0.55]{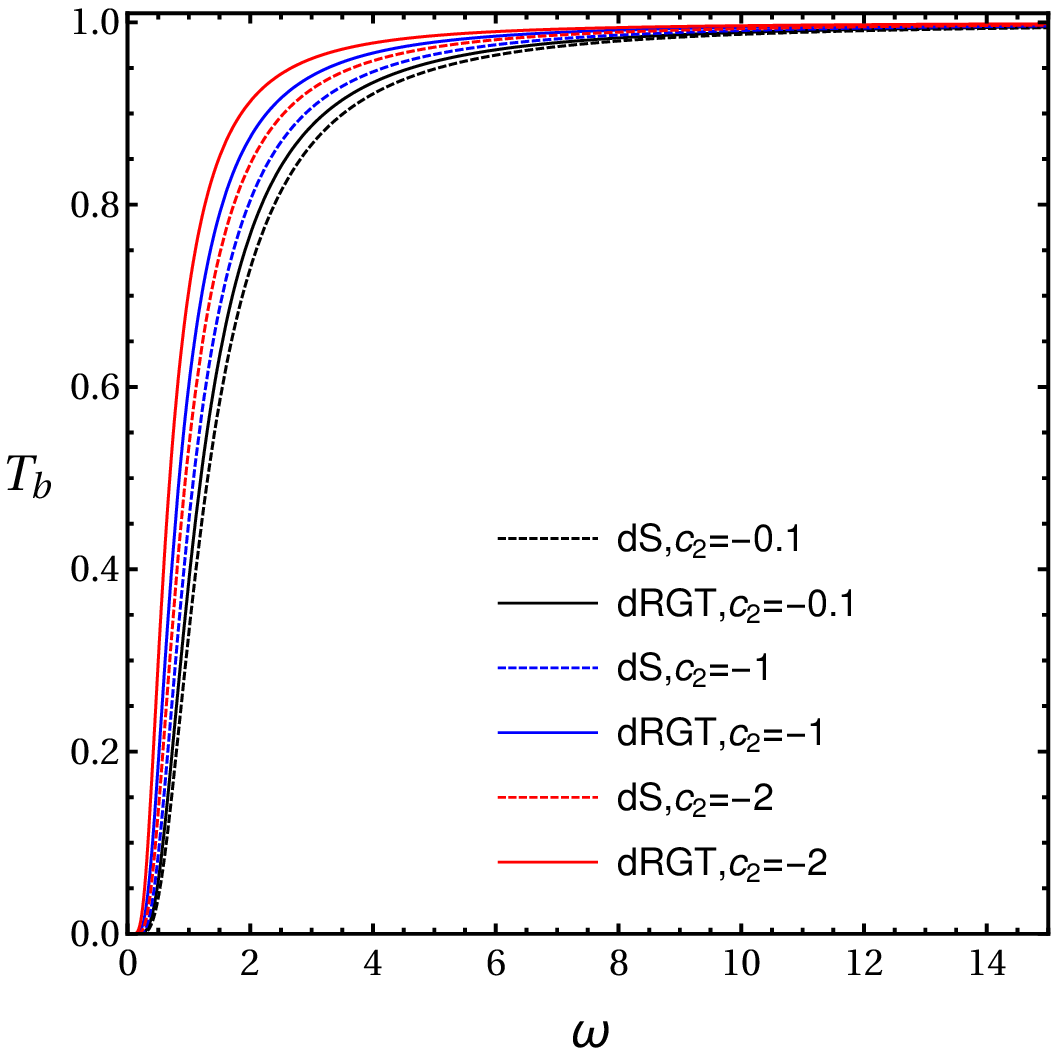}\qquad\qquad
\includegraphics[scale=0.55]{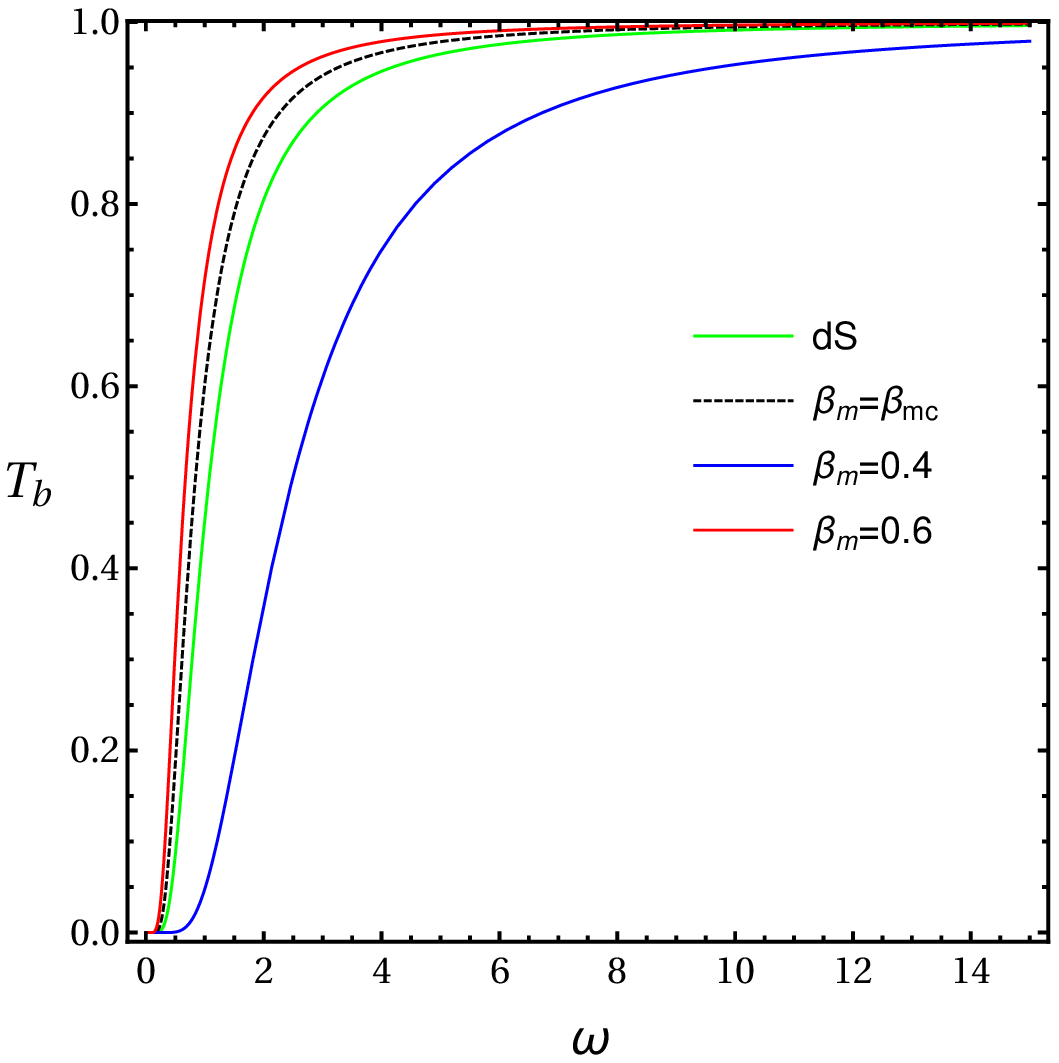}
\end{center}
{\caption{The left panel shows a comparison of the rigorous bound of greybody factor for the dRGT and the de Sitter black holes with $\ell$ = 0, $\tilde{M} = \alpha_{g} = 0.1$, while the parameter $\beta_m$ is chosen to have the same height of potential. The right panel shows the rigorous bound for $c_{2}$ = -1.}\label{TdRGT-fixc2}}
\end{figure}

For dRGT black holes, the potential is given by Eq. (\ref{poten mg}). Substituting this potential into Eq. (\ref{bound}), we obtain the rigorous bounds on the greybody factors
\begin{eqnarray}
T &\geq& \text{sech}^{2}\left[\frac{1}{2\omega c}\left\{\ell(\ell + 1)\left(\frac{1}{\tilde{r}_{H}} - \frac{1}{\tilde{R}_{H}}\right)\right.\right.\nonumber\\
  &&     + \tilde{M}\left(\frac{1}{\tilde{r}_{H}^{2}} - \frac{1}{\tilde{R}_{H}^{2}}\right) + 2\alpha_{g}c_{2}\left(\tilde{R}_{H} - \tilde{r}_{H}\right)\nonumber\\
  &&     \left.\left. - \alpha_{g}c_{1}\ln\left|\frac{\tilde{R}_{H}}{\tilde{r}_{H}}\right|\right\}\right].\label{TdRGT}
\end{eqnarray}
The rigorous bounds on the reflection probabilities are given by
\begin{eqnarray}
R &\leq& \tanh^{2}\left[\frac{1}{2\omega c}\left\{\ell(\ell + 1)\left(\frac{1}{\tilde{r}_{H}} - \frac{1}{\tilde{R}_{H}}\right)\right.\right.\nonumber\\
  &&     + \tilde{M}\left(\frac{1}{\tilde{r}_{H}^{2}} - \frac{1}{\tilde{R}_{H}^{2}}\right) + 2\alpha_{g}c_{2}\left(\tilde{R}_{H} - \tilde{r}_{H}\right)\nonumber\\
  &&     \left.\left. - \alpha_{g}c_{1}\ln\left|\frac{\tilde{R}_{H}}{\tilde{r}_{H}}\right|\right\}\right],
\end{eqnarray}
where, from Eqs. (\ref{rmg1}) and (\ref{rmg2}), $\tilde{r}_{H} = \tilde{r}_{dS1}$ and $\tilde{R}_{H} = \tilde{r}_{dS2}$ are given by
\begin{eqnarray}
\tilde{r}_{dS1} &=& \frac{-2}{\left(-2c_2\right)^{1/3}}\left[X^{1/2}\cos\left(\frac{1}{3}\sec^{-1}Y + \frac{\pi}{3}\right) + 1\right],\nonumber\\ \\
\tilde{r}_{dS2} &=& \frac{2}{\left(-2c_2\right)^{1/3}}\left[X^{1/2}\cos\left(\frac{1}{3}\sec^{-1}Y\right) - 1\right].
\end{eqnarray}
From Eq. (\ref{TdRGT}), we find that the rigorous bound on the greybody factor in massive gravity crucially depends on two parameters, $c_1$ and $c_2$, which determines how the structure of the graviton mass affects the bound. As we have discussed in Section \ref{sec:potential}, the parameter $c_1$ must be positive in order to have two horizons. Therefore, the last term in Eq. (\ref{TdRGT}) always provides the negative correction to the bound, so that, for the potentials with the same height, the bound from the dRGT black hole is always larger than the bound from the de Sitter black hole. Moreover, this behavior can be qualitatively expressed by analyzing the potential for both cases. From Fig. \ref{pot}, for the potentials with the same height, the potential from the dRGT case is always thinner than one from the de Sitter case. Therefore, the transmission amplitude for the dRGT case is always greater than one for the de Sitter case as seen in the left panel of Fig. \ref{TdRGT-fixc2}. As we have analyzed earlier, the height of the potential can be controlled by two parameters,  $\beta_m$ and $c_2$. Now let us figure out how the parameters affect the dRGT bound compared to the de Sitter bound.

By fixing $c_2$, one can see that the bound crucially depends on $|\tilde{R}_H - \tilde{r}_H|$, which is proportional to $1/\beta_m$. Therefore, one finds that the larger the value of $\beta_m$, the higher is the value of the bound. This can be seen explicitly by numerically plotting $T_b$ with various values of $\beta_m$ as illustrated in the right panel of Fig. \ref{TdRGT-fixc2}. Moreover, this behavior can also be seen by analyzing the potential. From Fig. \ref{pot}, we found that the larger the value of $\beta_m$, the lower is the peak of the potential. Therefore, one finds that the larger the value of $\beta_m$, the higher is the value of the bound.

\begin{figure}[h]
\begin{center}
\includegraphics[scale=0.6]{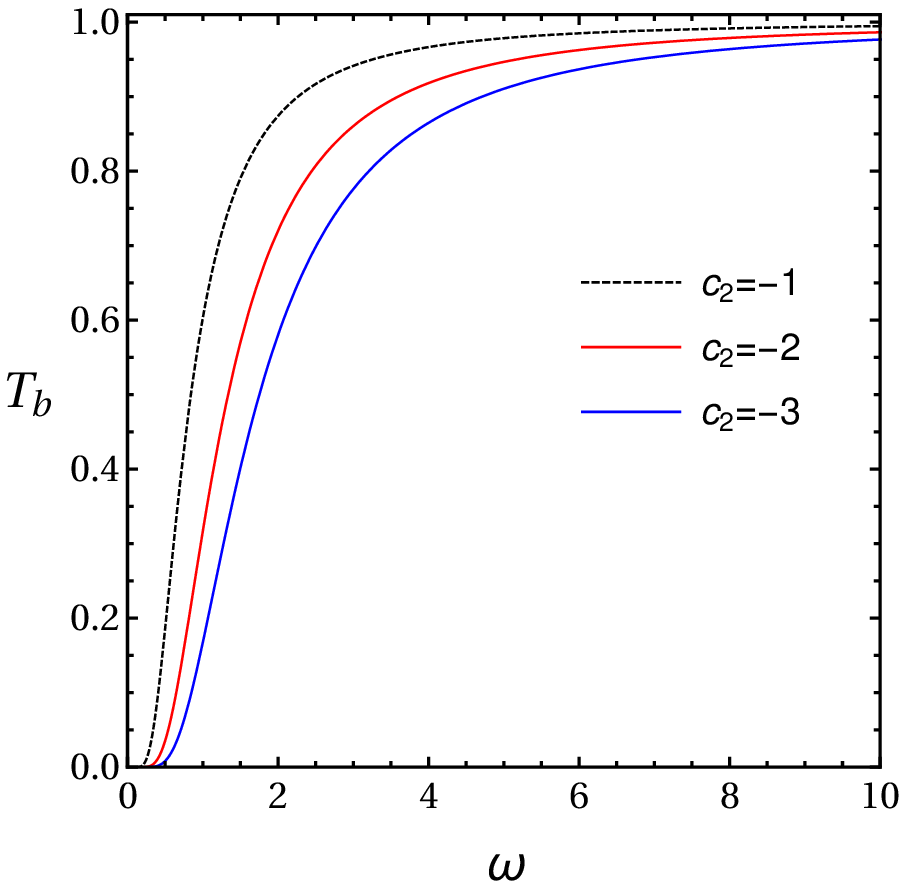}\qquad\qquad
\includegraphics[scale=0.6]{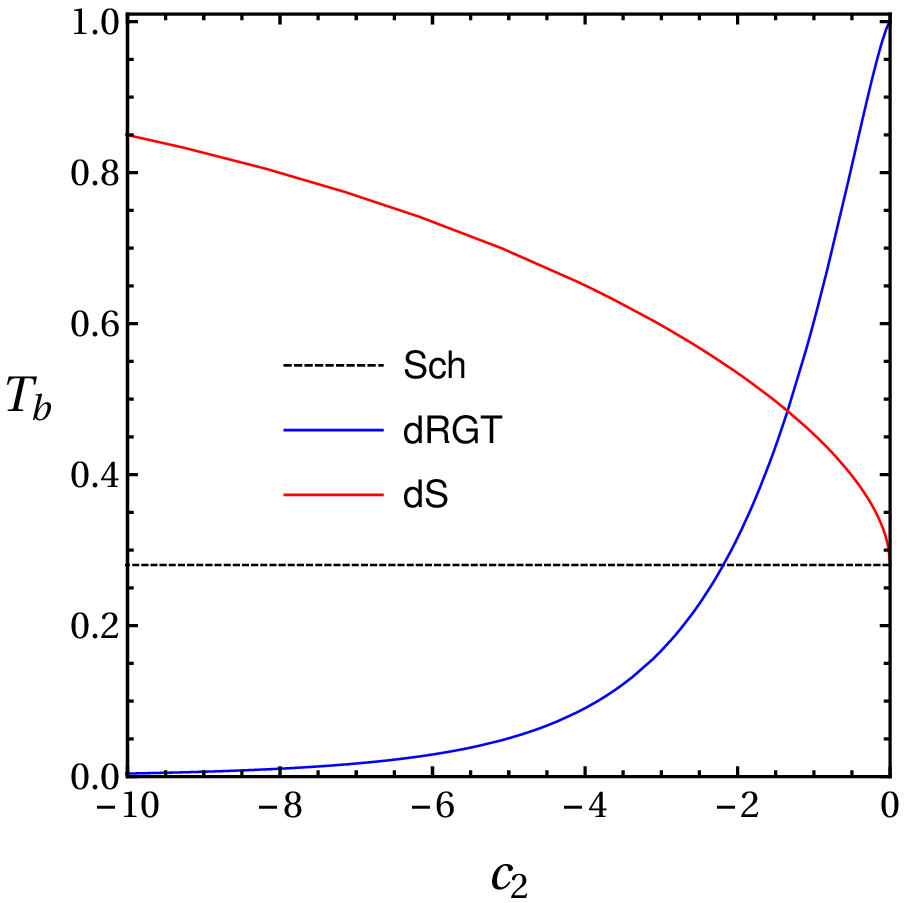}
\end{center}
{\caption{The left panel shows the effect of parameter $c_2$ on the rigorous bound on the greybody factor for a dRGT black hole with $\ell$ = 0, $\tilde{M}=\alpha_{g}$ = 0.1, and $\beta_{m}$ = 0.565375. The right panel shows a comparison of the rigorous bound on the greybody factor for a dRGT black hole with $\tilde{M}=\alpha_{g}= 0.1, \omega =1$, and $\beta_{m}$ = 0.565375.}\label{TdRGT-fixbetam}}
\end{figure}

In terms of fixing $\beta_m$, one can see that the maximum value of the potential decreases when $|c_2|$ decreases as such that the bound will increase when $|c_2|$ decreases as shown in the left panel of Fig. \ref{TdRGT-fixbetam}. Moreover, to compare the bound from the dRGT black hole to one from the de Sitter and the Schwarzschild black hole, we also plot the bound by fixing $\omega$ as seen in the right panel of Fig. \ref{TdRGT-fixbetam}. From this figure, we found that the bound from the dRGT black hole can be larger or smaller than ones from both the de Sitter and the Schwarzschild black holes, depending on $c_2$. On the other hand, the bound from the de Sitter black hole is always larger than the bound from the Schwarzschild black hole. Therefore, it is found that there is more room for the dRGT black hole to increase or decrease the greybody factor.

\section{\label{conclu}Conclusion}
In this paper, we obtain the gravitational potential from Schwarzschild black holes, de Sitter black holes, and dRGT black holes. We also derive the rigorous bound on the greybody factor for the de Sitter black hole and the dRGT black hole. It is found that the structure of potentials determines how much the rigorous bound on the greybody factor should be. Since $V_{\max}$ contributed from a de Sitter black hole is always less than one in a Schwarzschild black hole, the bound for a de Sitter black hole is greater than one for a Schwarzschild black hole. In case of potentials with the same height, the result shows that the bound from a dRGT black hole is always larger than the bound from a de Sitter black hole. Otherwise, the bound from a dRGT black hole can be larger or smaller than ones from both de Sitter and Schwarzschild black holes due to different effects of the parameter $c_{2}$ on de Sitter and dRGT spacetimes. Furthermore, we compare the greybody factor derived from the rigorous bound with the greybody factor derived from the matching technique. The results show that the greybody factor obtained from the rigorous bound is less than the one from the matching technique, which means that the rigorous bound is a true lower bound.

\begin{acknowledgments}
This project was funded by the Ratchadapisek Sompoch Endowment Fund, Chulalongkorn University (Sci-Super 2014-032), by a grant for the professional development of new academic staff from the Ratchadapisek Somphot Fund at Chulalongkorn University, by the Thailand Research Fund (TRF), and by the Office of the Higher Education Commission (OHEC), Faculty of Science, Chulalongkorn University (RSA5980038). PB was additionally supported by a scholarship from the Royal Government of Thailand. TN was also additionally supported by a scholarship from the Development and Promotion of Science and Technology Talents Project (DPST). PW was also additional supported by the Naresuan University Research Fund through grant No. R2559C235 and the ICTP through grant No. OEA-NET-01.
\end{acknowledgments}

\bibliography{apssamp}% Produces the bibliography via BibTeX.

\end{document}